\documentclass[twocolumn]{aastex63}
\usepackage{amsmath}
\usepackage{color}
\hypersetup{linkcolor=red,citecolor=blue}

\shortauthors{Zhong et al.}

\begin{document}
\title{\textbf{The magnetic origin of the mystery of rare H$\alpha$ Moreton waves}}

\correspondingauthor{Yao Chen}\email{yaochen@sdu.edu.cn}

\author[0000-0001-5483-6047]{Ze Zhong}
\affil{Center for Integrated Research on Space Science, Astronomy, and Physics, Institute of Frontier and Interdisciplinary Science, Shandong University, Qingdao 266237, China\\}
\affil{Institute of Space Sciences, Shandong University, Weihai 264209, China\\}

\author[0000-0001-6449-8838]{Yao Chen}
\affil{Center for Integrated Research on Space Science, Astronomy, and Physics, Institute of Frontier and Interdisciplinary Science, Shandong University, Qingdao 266237, China\\}
\affil{Institute of Space Sciences, Shandong University, Weihai 264209, China\\}

\author[0000-0002-9908-291X]{Y.W. Ni}
\affil{School of Astronomy and Space Science, Key Laboratory of Modern Astronomy and Astrophysics, Nanjing University, Nanjing 210023, China\\}

\author[0000-0002-7289-642X]{P. F. Chen}
\affil{School of Astronomy and Space Science, Key Laboratory of Modern Astronomy and Astrophysics, Nanjing University, Nanjing 210023, China\\}

\author[0000-0002-2734-8969]{Ruisheng Zheng}
\affil{Institute of Space Sciences, Shandong University, Weihai 264209, China\\}
\affil{Center for Integrated Research on Space Science, Astronomy, and Physics, Institute of Frontier and Interdisciplinary Science, Shandong University, Qingdao 266237, China\\}

\author[0000-0003-1034-5857]{Xiangliang Kong}
\affil{Institute of Space Sciences, Shandong University, Weihai 264209, China\\}
\affil{Center for Integrated Research on Space Science, Astronomy, and Physics, Institute of Frontier and Interdisciplinary Science, Shandong University, Qingdao 266237, China\\}

\author[0000-0001-7693-4908]{Chuan Li}
\affil{School of Astronomy and Space Science, Key Laboratory of Modern Astronomy and Astrophysics, Nanjing University, Nanjing 210023, China\\}
\affil{Institute of Science and Technology for Deep Space Exploration, Suzhou Campus, Nanjing University, Suzhou 215163, China\\}

\begin{abstract}
Over the past three decades, a lot of coronal fast-mode waves were detected by space missions, but their counterparts in the chromosphere, called the Moreton waves, were rarely captured. How this happens remains a mystery. Here, to shed light on this problem, we investigate the photospheric vector magnetograms of the Moreton wave events associated with M- and X-class solar flares in 2010--2023. The H$\alpha$ data are taken with the Global Oscillation Network Group (GONG) and the Chinese H$\alpha$ Solar Explorer (CHASE). Our statistical results show that more than 80\% of the events occur at the edge of active regions and propagate non-radially due to asymmetric magnetic fields above the flares. According to the reconstructed magnetic field and atmospheric model, Moreton waves propagate in the direction along which the horizontal fast-mode wave speed drops the fastest. The result supports that the inclined magnetic configuration of the eruption is crucial to generate Moreton waves, even for X-class flares. It may explain the low occurrence rate of Moreton waves and why some X-class flares accompanied with coronal mass ejections (CMEs) do not generate Moreton waves.
\end{abstract}

\keywords{Solar coronal waves (1995) --- Solar chromosphere (1479) --- Solar activity (1475) --- Solar Magnetic fields (1503)}

\section{Introduction} \label{sec:intro}
Solar eruptions often cause disturbances propagating through the corona on a global scale (e.g., see reviews by \citealt{2015Warmuth,2016Chen}) with various effects in the chromosphere, transition region, and corona. One of the most spectacular disturbances is Moreton waves, which are typically observed with the H$\alpha$ spectral line \citep{1960Moreton1,1960Moreton2}. Observations show that the characteristics of Moreton waves are bright arcs in the H$\alpha$ blue wing (or dark arcs in the red wing) \citep{1961Athay}, indicating that the corresponding chromospheric materials are pushed downward. Moreton waves propagate in a narrow angular span with speeds ranging from $\sim$300 km s$^{-1}$ to more than 2000 km s$^{-1}$ \citep{1960Moreton1,1971Smith,2011Zhang}. Another type of global waves are coronal waves. They have been observed in a wide range of extreme ultraviolet (EUV) wavelengths \citep[e.g.,][]{2009Thompson,2012Shen,2013Nitta,2014Muhr,2019Shen,2022Zheng}, soft X-rays \citep[e.g.,][]{2002Khan,2005Warmuth,2009Attrill}, and radio wavelengths \citep[e.g.,][]{2005Vrsnak,2005White,2014Chen,2023Lv}. Coronal waves have various shapes, with speeds varying from tens km s$^{-1}$ to 1500 km s$^{-1}$ \citep{2009Thompson,2013Nitta}, and can be used to estimate the magnetic field intensity in the corona \citep{2024Nakariakov}.

The widely accepted explanation for Moreton waves is that they are the chromospheric response to the fast-mode magneto-hydrodynamic (MHD) shock propagating in the corona \citep{1968Uchida}, though \citet{2002Chen} proposed that the fast-mode shock is due to coronal mass ejections (CMEs) rather than the pressure pulse in solar flares. On the contrary, there are several competing interpretations on the nature of coronal waves, such as MHD fast-mode waves \citep{2000Vrsnak}, MHD slow-mode soliton waves \citep{2007Wills}, non-wave models \citep{2007Attrill,2008Delannee}, a hybrid model \citep{2002Chen,2005Chenb}, etc. According to the hybrid model, there are two types of coronal EUV waves in a single event, a faster one and a slower one. The faster one is generally sharp and bright, and is considered as a fast-mode wave or shock wave. Sometimes, faster EUV waves are cospatial with H$\alpha$ Moreton waves and are related to type II radio bursts \citep{2004Warmuth2}; In contrast, the slower one is generally diffuse, and is interpreted as an apparent propagation due to successive magnetic fieldline stretching \citep{2002Chen}. 

Despite hot debates on the nature of the slower EUV waves, it has been well established that the faster EUV waves are the coronal counterparts of chromospheric Moreton waves. When the coronal fast-mode wave sweeps the chromosphere, pushing the chromospheric material downward, it generates an apparent propagation of the H$\alpha$ Moreton wave. However, while hundreds of coronal fast-mode waves have been observed \citep{2013Nitta} since the Solar Dynamics Observatory \citep[SDO;][]{2012Pesnell} was launched in 2010, only ten Moreton waves have been reported \citep{2012Asai,2015Admiranto,2016Francile,2019Long,2019Cabezas,2020Wang,2023Zheng}. Part of the reasons is that the chromosphere is several orders of magnitude denser than the corona so that the chromospheric imprints of coronal shock waves are much weaker or even invisible. Still such a huge gap remains a mystery over past decades.

Several studies have attempted to interpret the discrepancy. One possibility is that the Moreton wave is visible only when the CME experiences an overexpansion in the early stage \citep{2016Vrsnak,2018Krause}. Another possibility is that the CME eruption is significantly inclined. In this case, a strong perturbation sweeps the chromosphere on the side of the inclined eruption even when the eruption is relatively weaker \citep{2016Vrsnak}. \citet{2023Zheng} analyzed three H$\alpha$ Moreton waves associated with B-, M-, and X-class flares, and demonstrated that highly inclined eruptions are critical to generate Moreton waves. However, they did not consider the magnetic field configuration which is important in channeling the eruption. 

In this paper, we investigate 66 Moreton wave events associated with M- and X-class flares observed during the SDO era, with a much larger sample than previous studies. Our purpose is to reveal the magnetic field configuration and answer the question why Moreton waves are so rare. We introduce the observations and data analysis in Section \ref{s:obs}, and present the results in Section \ref{s:results}. The summary and discussion are given in Section \ref{sec:discussion}.

\section{Observations and Data Analysis}\label{s:obs}
Using the H$\alpha$ data, we found 66 Moreton wave events from 2010 April to 2023 December. All Moreton waves are monitored by the Global Oscillation Network Group \citep[GONG;][]{2011Harvey} and six of them are detected by the Chinese H$\alpha$ Solar Explorer \citep[CHASE;][]{2022Li}. Note that in this paper we select the Moreton wave events where the wave signature is visible in 3 consecutive frames of images. The time cadence of GONG is 60 s and the pixel size is 1$\arcsec$. CHASE obtains full-disk spectroscopic H$\alpha$ (6559.7--6565.9 {\AA}) data within a cadence of $\sim$73 s and a spatial resolution of 1\farcs04 per pixel in the binning mode \citep{2022Qiu}. By fitting the H$\alpha$ spectral profiles, we can derive the dopplergrams of the full solar disk. With high spatial resolution observations from the Atmospheric Imaging Assembly \citep[AIA;][]{2012Lemen} on board SDO and the Extreme UltraViolet Imager (EUVI) on board the Solar Terrestrial Relations Observatory \citep[STEREO;][]{2008Kaiser}, 302 coronal EUV wave events are detected during the same period, about 5 times more than Moreton waves. All Moreton waves have corresponding EUV waves.

\begin{figure*}
\centering
\includegraphics[width=1.0\textwidth]{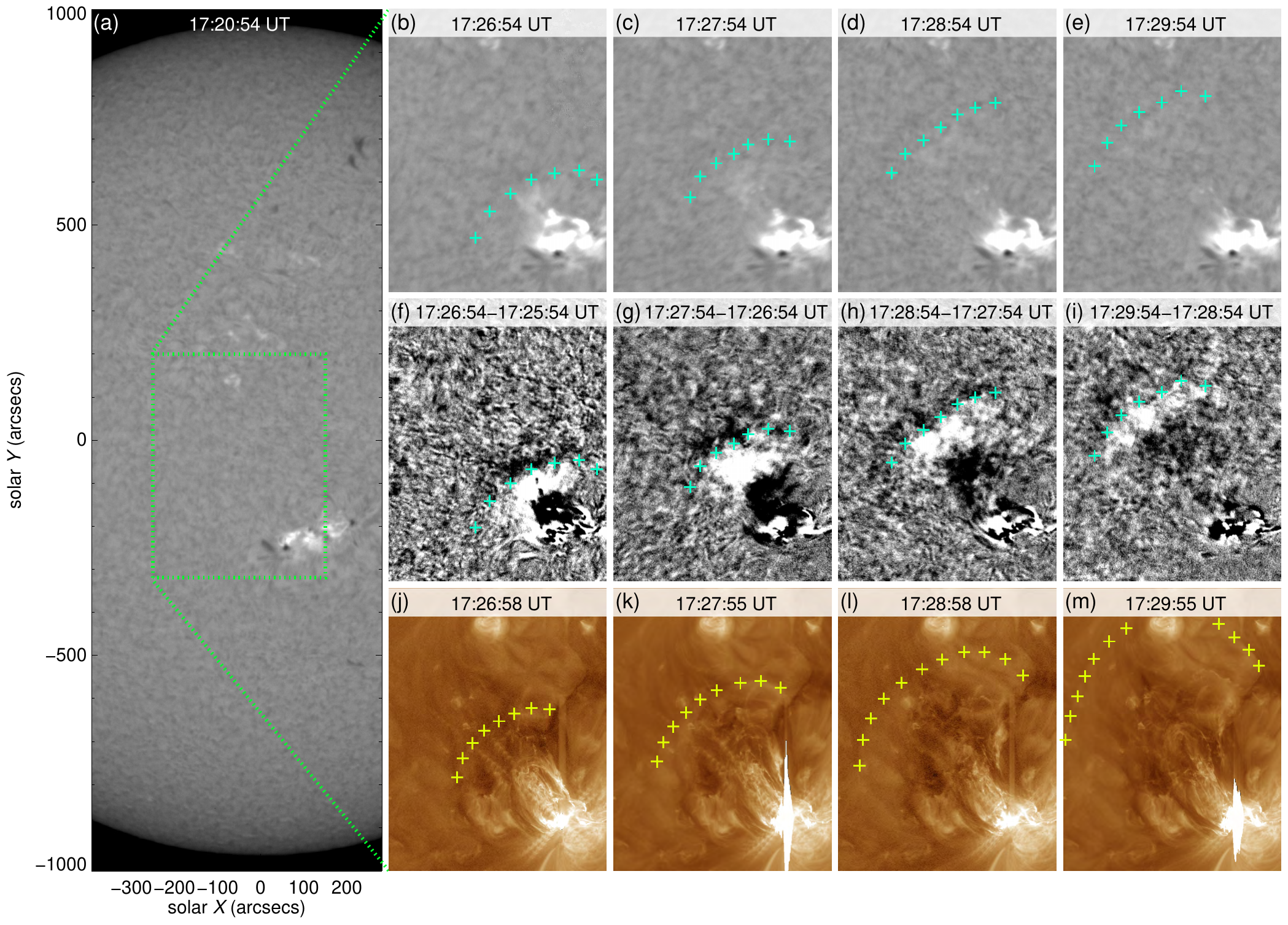}
\caption{Propagation of the Moreton wave monitored by GONG and the coronal wave observed by AIA.
(a) Full overview of the flare occurred at 17:20:54 UT on 2011 February 14. 
(b--e) The sequence of H$\alpha$ images from 17:26:54 to 17:29:54 UT with a zoomed-in view of the region corresponding to the green dotted box in panel (a). Cyan crosses depict the front of the Moreton wave.
(f--i) Evolution of the Moreton wave with H$\alpha$ running-difference images. The cyan crosses are the same as that in panels (b)--(e).
(j--m) Time sequence data of AIA 193 {\AA} images show the morphological evolution of the coronal wave. Yellow crosses mark the wave front.
(An animation of this figure is available. It shows the evolution of the Moreton wave monitored by GONG and the associated coronal EUV wave observed by AIA from 17:21:54 UT to 17:34:54 UT. The duration of the animation is 2.8 seconds.)
}
\label{fig01}
\end{figure*}

We present two examples to illustrate our selection and analysis process. At 17:20:54 UT on 2011 February 14, an M2.2-class flare occurred in the active region (AR) 11158 (S20W04; Figure \ref{fig01}(a)), which was registered by the Geostationary Operational Environmental Satellites (GOES). Five minutes later, GONG monitored a Moreton wave in the H$\alpha$ line center from 17:25:54 UT to 17:29:54 UT. It mainly traveled northeast from the flare region. Figures \ref{fig01}(b)--(e) and (f)--(i) show four H$\alpha$ images and their running-difference images, with the wave front marked by the cyan crosses. The average propagation speed is about 625.33 $\pm$ 53.32 km s$^{-1}$ during the 4 minutes appearance of the Moreton wave. We also used AIA 193 {\AA} images to compare the H$\alpha$ Moreton wave with the EUV wave as shown in Figures \ref{fig01} (j)--(m). We labeled a series of sharp wave fronts from 17:26:58 UT to 17:29:55 UT with yellow crosses. The EUV wave is ahead of the Moreton wave, with a wider wave front propagating at a speed of $\sim$635.33 $\pm$ 14.14 km s$^{-1}$. 

\begin{figure*}
\centering
\includegraphics[width=1.0\textwidth]{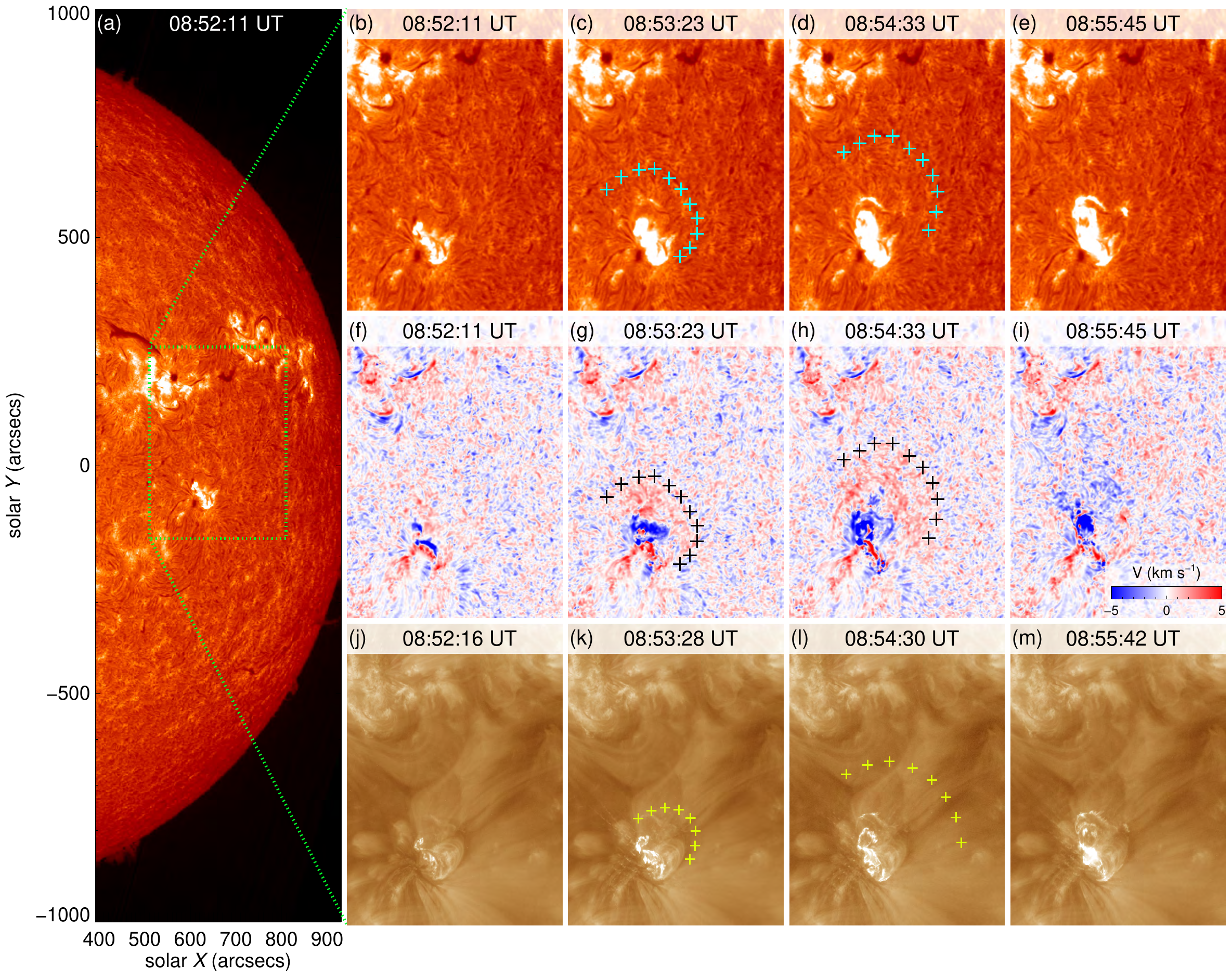}
\caption{Evolution of the Moreton wave observed by CHASE and the coronal wave detected by AIA.
(a) Full overview of the Moreton wave occurred at 08:52:11 UT on 2023 May 11. 
(b--e) The sequence of images in the H$\alpha$ line center from 08:52:11 to 08:55:45 UT with a zoomed-in view of the region corresponding to the green dotted box in panel (a). Cyan crosses depict the front of the Moreton wave.
(f--i) Propagation of the Moreton wave with H$\alpha$ dopplergrams derived by subtracting the blue wing (-0.5 {\AA}) from the red wing (+0.5 {\AA}). The black crosses are the same as that in panels (b)--(e).
(j--m) Time sequence data of AIA 193 {\AA} images display the evolution of the coronal wave. Yellow crosses outline the wave front.
(An animation of this figure is available. It shows the evolution of the Moreton wave detected by CHASE and the associated coronal EUV wave observed by AIA from 08:52:11 UT to 09:06:23 UT. The duration of the animation is 2.6 seconds.)
}
\label{fig02}
\end{figure*}

Figure \ref{fig02} shows another example of the Moreton wave observed by CHASE on 2023 May 11 (S06W41; Figure \ref{fig02}(a)). At 08:47:11 UT, a GOES M2.2-class flare occurred in the AR 13294. CHASE captured a Moreton wave in the H$\alpha$ line center, red and blue wings after five minutes. Figures \ref{fig02}(b)--(e) and (f)--(i) show four H$\alpha$ line center images and the derived dopplergrams, with the wave front (cyan crosses) observed from 08:53:23 to 08:54:33 UT. The velocity of the Moreton wave is about 422.57 $\pm$ 27.23 km s$^{-1}$ during 3 minutes. The corresponding EUV wave is also observed in the AIA 193 {\AA} data (yellow crosses) with a speed of $\sim$551.17 $\pm$ 19.68 km s$^{-1}$, as shown in Figures \ref{fig02}(j)--(m).

All these eruption events are also observed by the Helioseismic and Magnetic Imager \citep[HMI;][]{2012Scherrer} on board SDO. The photospheric vector magnetograms are inverted through fitting the observed Stokes profiles using the Very Fast Inversion of the Stokes Vector \citep[VFISV;][]{2011Borrero}, which is provided by the HMI pipeline (\citealt{2014Hoeksema}). The data cover the full solar disk with a cadence of 12 minutes and a pixel size of $0.505\arcsec$ \citep{2012Schou}. We also require the heliocentric angle of the source active regions to be less than 75$^\circ$ for a proper analysis on the magnetic field configuration. Then, we project the magnetogram to the heliographic coordinate using the method proposed by \citet{1990Gary} and reconstruct the potential field from the normal component of the vector magnetograms using the Green's function method \citep{1977Chiu}. The information of the 66 events of Moreton waves is exhibited in Table \ref{tab1}.

\section{Results}\label{s:results}
\subsection{Location of the Flare on the Photospheric Magnetograms}\label{ss:fe}
Flare ribbons in the chromosphere outline the footpoints of flare loops. They often brighten impulsively in the UV wavelength before flare loops are visible in the EUV passband. We divide the 66 wave events into three groups according to the relative location of the flare ribbons on the magnetograms. Figures \ref{fig03}--\ref{fig05} show the mapping of the flare ribbons on the HMI line-of-sight magnetograms for all events in the three groups, respectively. The flare ribbons in 1600 {\AA} exhibit complex pattern of motions, including the perpendicular separation (e.g. Figure \ref{fig03}(a)), elongation parallel to the polarity inversion line (e.g. Figure \ref{fig04}(ac)), and other irregular motions. The duration of the ribbon motion is 15 to 30 minutes. The main characteristics of three groups are summarized as follows:

\begin{enumerate}
\item Group--\uppercase\expandafter{\romannumeral1}:
The flare ribbons appear at the edge of the source active region, and the center of the active region occupies one (e.g. Figure \ref{fig03}(a)) or multiple (e.g. Figure \ref{fig03}(d)) main magnetic polarities, forming an open field configuration. The total magnetic flux of the active region shows a strong imbalance. There are 24 events belonging to Group--\uppercase\expandafter{\romannumeral1}. Several Moreton waves in this group are homologous events, e.g., the Moreton waves, which occurred on 2011 October 1 and 2 (Figures \ref{fig03} (a)--(b)), 2014 March 28, 29 and 30 (Figures \ref{fig03} (h)--(k)), 2021 May 22 and 23 (Figures \ref{fig03} (m)--(o)), 2022 August 15 and 16 (Figures \ref{fig03} (q)--(r)), and 2023 September 19, 20 and 21 (Figures \ref{fig03} (v)--(x)), originated from ARs 11305, 12017, 12824, 13078, and 13435, respectively.

\item Group--\uppercase\expandafter{\romannumeral2}:
The flare ribbons also appear at the edge of the source active region, but the photospheric magnetograms display one (e.g., Figure \ref{fig04}(b)) or two (e.g., Figure \ref{fig04}(a)) pairs of magnetic polarities at the center of the active region, forming large-scale strong magnetic loops. As shown in Figure \ref{fig04}, there are 29 events belonging to Group--\uppercase\expandafter{\romannumeral2}. There are several homologous events in this group, e.g., the Moreton waves, which occurred on 2011 September 24 and 25 (Figures \ref{fig04} (d)--(f)), 2013 November 8 and 10 (Figures \ref{fig04} (o)--(p)), 2015 November 4 (Figures \ref{fig04} (q)--(r)), 2017 April 1 and 2 (Figures \ref{fig04} (t)--(u)) and 2022 March 30 and 31 (Figures \ref{fig04} (w)--(x)), originated from ARs 11302, 11890, 12445, 12644, and 12975, respectively.

\item Group--\uppercase\expandafter{\romannumeral3}: 
The flare ribbons are located close to the center of the active region. Half of events exhibit a typical two-ribbon property (e.g., Figure \ref{fig05}(a)) demonstrating the configuration of standard flare model. For the morphology of the photospheric magnetograms, there are no unified characteristics. 

\end{enumerate}
It is noted that Moreton waves of Groups \uppercase\expandafter{\romannumeral1} and \uppercase\expandafter{\romannumeral2} occupy 80.3\% of our sample.

\subsection{Magnetic Configuration Generating the Moreton Wave}\label{ss:mgc}
 We reconstruct the corresponding coronal magnetic field to further investigate the magnetic configuration responsible for the generation of Moreton waves. For Group--\uppercase\expandafter{\romannumeral1}, we take the Moreton wave on 2011 October 1 in the AR 11305 as an example. The associated M1.2-class flare began at 09:40 UT and the Moreton wave was visible from 09:45:14 to 09:54:14 UT, propagating southward with an arc shape (Figure \ref{fig06}(a) yellow plus signs). Figure \ref{fig06}(b) shows that the two bright flare ribbons are respectively situated on the weaker positive polarity P1 in the south and the southern edge of the much stronger negative polarity N1 in the north due to the significantly uneven magnetic flux. As indicated by the extrapolated coronal magnetic field in Figures \ref{fig06}(c) and \ref{fig06}(d), a bundle of open field lines is rooted to the north of the closed field lines above the flare ribbons. Therefore, the open field lines act as a wall, causing the closed magnetic loops inclined toward the weaker field, and leading the eruption to be biased toward the south with weaker magnetic field. 

 We also reconstruct the coronal magnetic field for other events in Group--\uppercase\expandafter{\romannumeral1}, and the results are similar. For all events, the distribution of the background magnetic field is strongly uneven, causing the Moreton wave to propagate along the direction with weaker field. Meanwhile, we compare the results of the coronal magnetic field using potential field extrapolation methods \citep{1977Chiu,1981Alissandrakis} and nonlinear force-free field (NLFFF) extrapolation method \citep{2004Wiegelmann}. The NLFFF model can reconstruct magnetic sheared structures above the polarity inversion line in flare regions. For the background magnetic field, different extrapolation methods provide similar results.

In Group--\uppercase\expandafter{\romannumeral2}, we described the observational characteristics of the representative Moreton wave in Section \ref{s:obs}. Taking the 2011 February 14 Moreton wave event as an example, the flare appears at the northeast edge of the active region, as shown in Figures \ref{fig06}(e) and \ref{fig06}(f). The active region consists of two bipoles (denoted as P1/N1 and P2/N2 in Figure \ref{fig06}(f)) that form a quadrupolar configuration. To the north of the negative polarity N1, there emerge a few positive polarities P3. The interaction between them triggers the M2.2-class flare ultimately \citep{2012Sun}. Based on the extrapolated magnetic field shown in Figures \ref{fig06}(g) and \ref{fig06}(h), we find that the two flare ribbons are located on the minority positive polarity P3 and the northern edge of the negative polarity N1. Similar to the Group--\uppercase\expandafter{\romannumeral1} example, large-scale strong magnetic loops stand to the south of the flare loop, acting as a wall to make the eruption channeled to the north. As a result, the Moreton wave is seen to propagate to the north.

The overlying fields have the inclined configuration above the flare for the first two groups. For Group--\uppercase\expandafter{\romannumeral3}, except for the 2014 July 8 (Figure \ref{fig05}(g)) and 2022 October 2 (Figure \ref{fig05}(k)) events, the magnetic configurations of other events are overlaid with symmetric magnetic loops generated by the bipolar field. The chromospheric response presents an arc-shaped front of all waves, which suggests the eruption tilt to one side. We investigate two events to explore the reason of non-radial eruption. The X1.6-class flare on 2014 September 11 (Figure \ref{fig05}(h)) displays an asymmetric feature for a two-ribbon flare. \citet{2021Kilpua} simulated this event and revealed its eruption with an inclination. Another flare on 2011 August 4 (Figure \ref{fig05}(b)) shows a parallel two-ribbon property. \citet{2023Zhong} performed a data-driven simulation for this flare and reproduced the inclined eruption, which is exactly along the direction of the Moreton wave propagation. We find that the asymmetry between the footpoints of the flux rope leads to their non-radial eruption, then to generate the Moreton waves, even when the overlying fields seem to be symmetric.

\subsection{Propagation Direction of Moreton Waves}\label{ss:fp}
We extrapolate the coronal magnetic field for all events and confirm that Moreton waves propagate along the direction toward weaker magnetic field, similar to the previous studies \citep{2011Zhang,2019Cabezas}, but with a larger sample. The propagation speed of the fast-mode shock differs by an order of magnitude between the corona and the chromosphere. Can we predict the propagation direction of the Moreton wave according to the distribution of the fast-mode wave speed? To do this, we calculate the distribution of the fast-mode magnetoacoustic speed using an atmospheric model combined with the extrapolated magnetic field. We use a stratified profile to mimic the solar atmosphere from the chromosphere to the corona. The temperature distribution is set as

\begin{equation}
T= T_{_{\rm ch}}+(T_{_{\rm co}}-T_{_{\rm ch}})({\rm tanh}(\frac{h-h_{_{\rm tr}}}{w_{\rm tr}})+1)/2,
\end{equation}
where $T_{_{\rm ch}}=8000 \rm \ K$ is the chromospheric temperature, $T_{_{\rm co}}=2.5 \rm \ MK$ is the coronal temperature, $h_{_{\rm tr}}=5 \rm \ Mm$ is the height of the initial transition region, $w_{_{\rm tr}}=0.5 \ \rm Mm$ is the thickness of the initial transition region. The density is obtained by solving the hydrostatic equilibrium equation, $\emph{d} (\rho T) / \emph{d} h = -\rho g$. 

Taking the 2011 October 1 M1.2-class flare as an example, we determine the location of the core (the red circles in Figures \ref{fig07}(a) and \ref{fig07}(b), which display the magnetic field distribution from two perspectives) and acquire the distribution of surrounding fast-mode magnetoacoustic speed in the height of the bottom corona. We find that the direction along which the horizontal fast-mode wave speed drops the fastest, as indicated by the red arrow, coincides well with the propagation direction of the Moreton wave as shown in the animation. We also apply the model to the M2.2-class flare on 2011 February 14 (Figures \ref{fig07}(c) and \ref{fig07}(d)) and confirm that Moreton waves propagate in the direction along which the speed of the horizontal fast-mode wave in the bottom corona drops the fastest.

\section{Summary and Discussion} \label{sec:discussion}
In this paper, we investigated the magnetic configuration of 66 Moreton wave events associated with M- and X-class flares observed by GONG and CHASE from 2010 to 2023. The photospheric magnetograms show that 80.3\% of events occur at the edges of active regions. Further analysis reveals that the magnetic configurations of all events have the inclined or asymmetric structures, leading to non-radial eruptions. Combined with the atmospheric model, we found Moreton waves propagate along the direction with the fastest decrease of the horizontal fast-mode wave speed.

\subsection{Can Radial Eruption Generate Moreton Waves?}\label{ss:can}
The magnetic configurations of the 66 events demonstrate that both the inhomogeneity of the overlying fields and the asymmetry of flux rope footpoints can lead to non-radial eruptions. It leaves a question on whether Moreton waves can generate in radial eruptions. To address this issue, we first analyze an X1.4-class flare near the solar center associated with a fast halo CME as an example of radial eruptions. The flare started at 15:37 UT on 2012 July 12, and then peaked at 16:49 UT. The circular front of the EUV wave appeared at 16:15 UT and seen clearly at 16:25 UT on the AIA 193 {\AA} image (cyan dots in Figure \ref{fig08}(a)). In Figure \ref{fig08}(b) with the same field of view, the wave has no counterpart in the AIA 304 {\AA} image. It may indicate that this eruption is hard to disturb the transition region and the chromosphere. According to the flare ribbons superimposed on the magnetogram in Figure \ref{fig08}(c), the ejecta is confined by strong background magnetic field in every direction, and is difficult to be deflected.

Then, we investigate a radial eruption that occurred above the solar limb. The eruption produced an X8.2-class flare, ranking the second largest one in solar cycle 24 \citep{2018Hou}. After a tear-drop-shaped structure escaped from the solar western limb (Figure \ref{fig08}(d)), a coronal wave was visible from 15:56 UT to 16:06 UT in the AIA 304 {\AA} images. We measure the speed of the coronal wave along the cyan line, which is about 677.16 $\pm$ 24.95 km s$^{-1}$ according to the time--distance diagram in Figure \ref{fig08}(e). However, the wave has no response in the H$\alpha$ diagram (Figure \ref{fig08}(f)). The two examples demonstrate that it is difficult to generate H$\alpha$ Moreton waves for radial eruptions, even for X-class flares. 

\subsection{Possible Reasons for the Low Occurrence Rate of Moreton Waves}\label{ss:is}
In our statistics, although the number of Moreton waves is larger than in previous studies, there is a huge gap compared to coronal fast-mode waves. One straightforward reason is that the chromosphere is much denser than the corona, hence the chromosphere can hardly be disturbed unless the eruption is strong enough. Other factors should be explored further. Previous studies considered that overexpansion and highly inclined configuration of pre-eruption are important to generate Moreton waves. We re-examined the independent role of the overexpansion. From the perspective of the observations, GONG did not capture an observable Moreton wave caused by overexpansion in the radial eruption, which are only seen in the symmetric MHD simulations \citep{2005Chen,2016Vrsnak}. In contrast, \citet{2019Long} found that the density compression ratios of the observed Moreton waves are much lower than those predicted by \citet{2016Vrsnak} from MHD simulations. Therefore, these results seem not to support the necessity and sufficiency of the overexpansion in generating Moreton waves. 

\begin{figure*}
\centering
\includegraphics[width=1.0\textwidth]{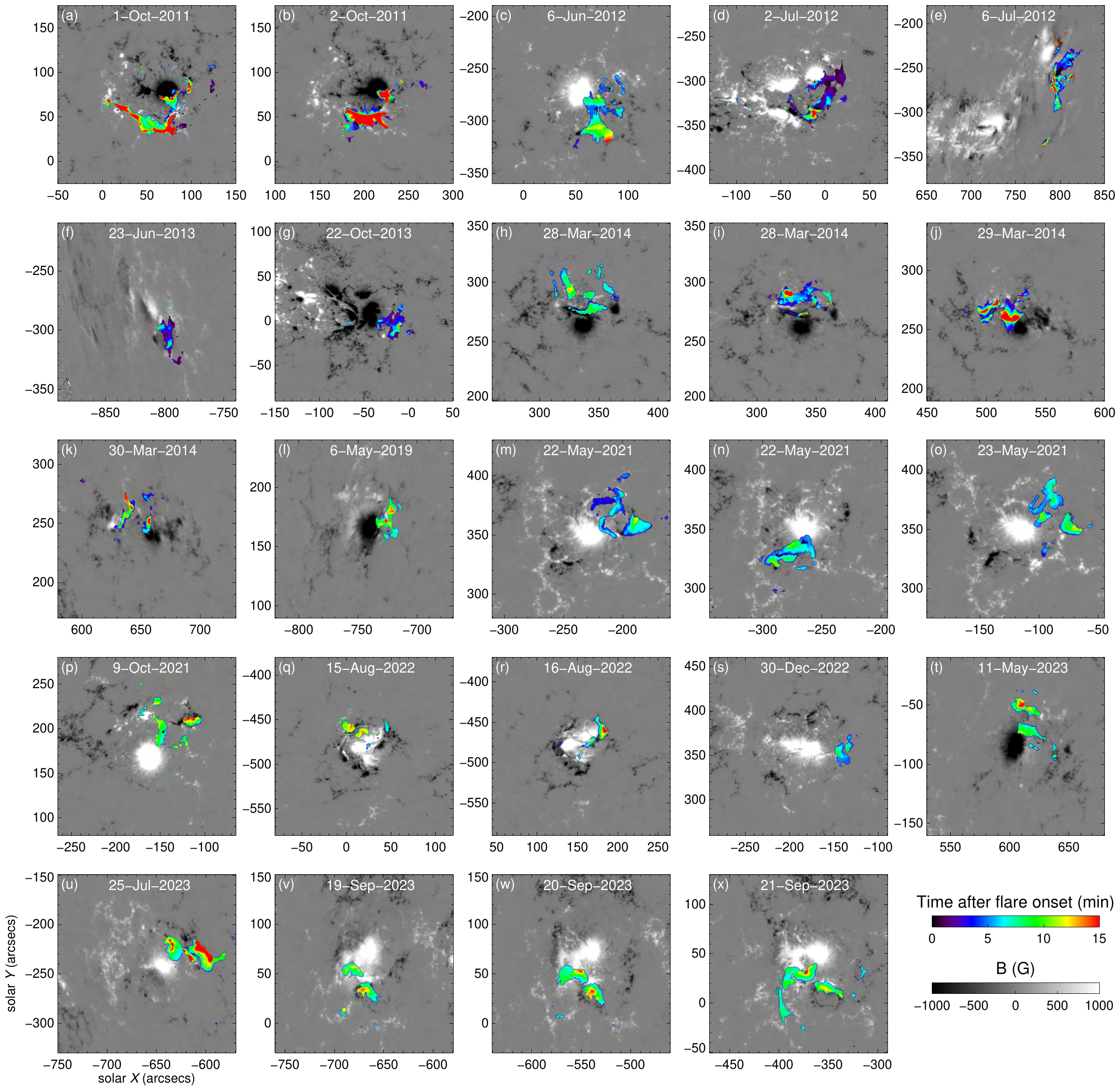}
\caption{Group--\uppercase\expandafter{\romannumeral1} of HMI line-of-sight photospheric magnetograms (grayscale) includes 24 Moreton wave events. All magnetograms display one (e.g. panel (a)) or multiple (e.g. panel (d)) main magnetic polarities in the center of the active region, with opposite polarities in the periphery. The newly brightened flare ribbons are superimposed on the magnetogram, with the time lapse given by the rainbow color bar. Flare ribbons of all events are located at the edge of the active region.
} 
\label{fig03}
\end{figure*}

\begin{figure*}
\centering
\includegraphics[width=1.0\textwidth]{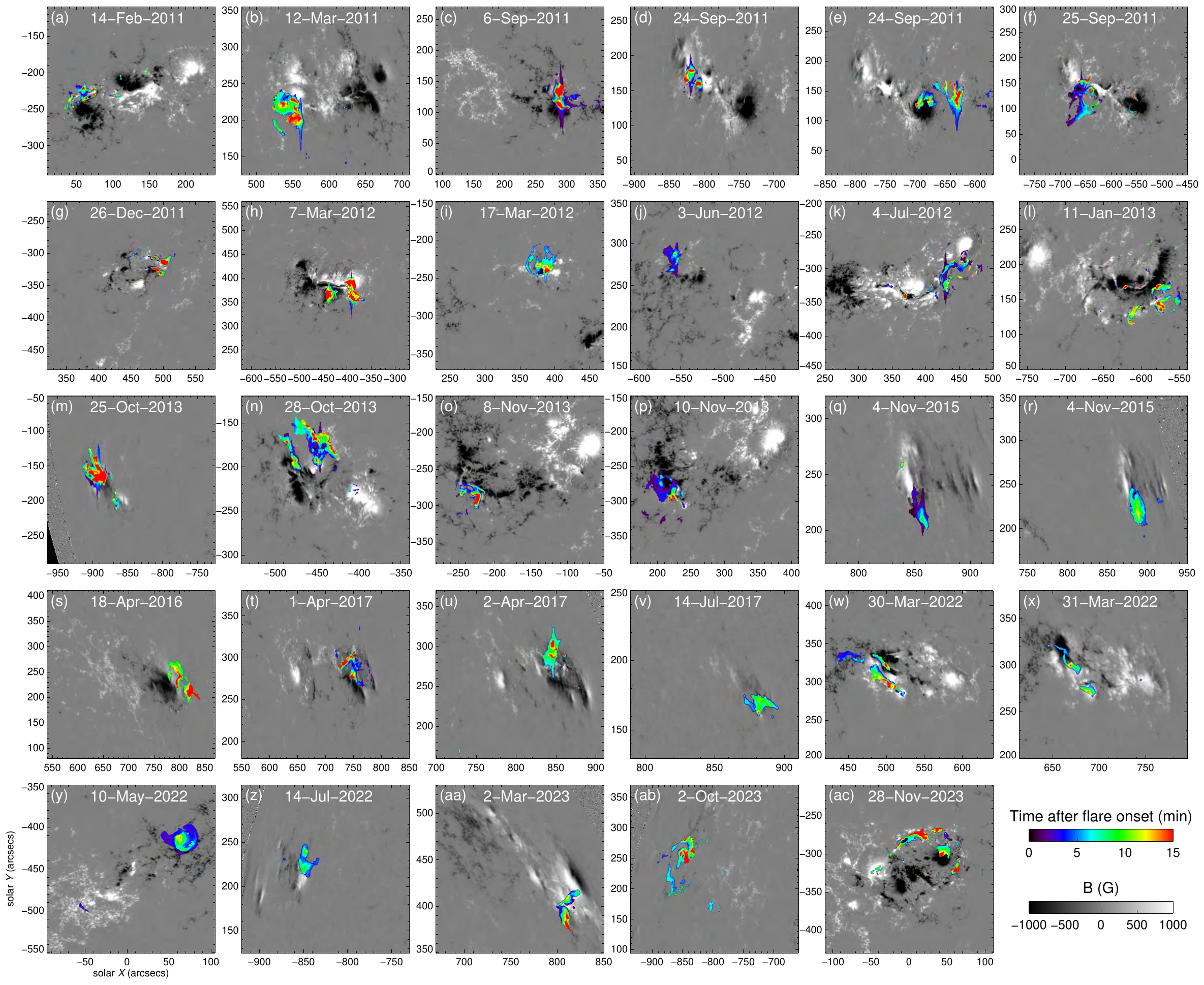}
\caption{Similar to Figure \ref{fig03} but for Group--\uppercase\expandafter{\romannumeral2} containing 29 Moreton wave events. All magnetograms show one (e.g. panel (b)) or two (e.g. panel (a)) pairs of main magnetic polarities in the center of the active region.
} 
\label{fig04}
\end{figure*}

\begin{figure*}
\centering
\includegraphics[width=1.0\textwidth]{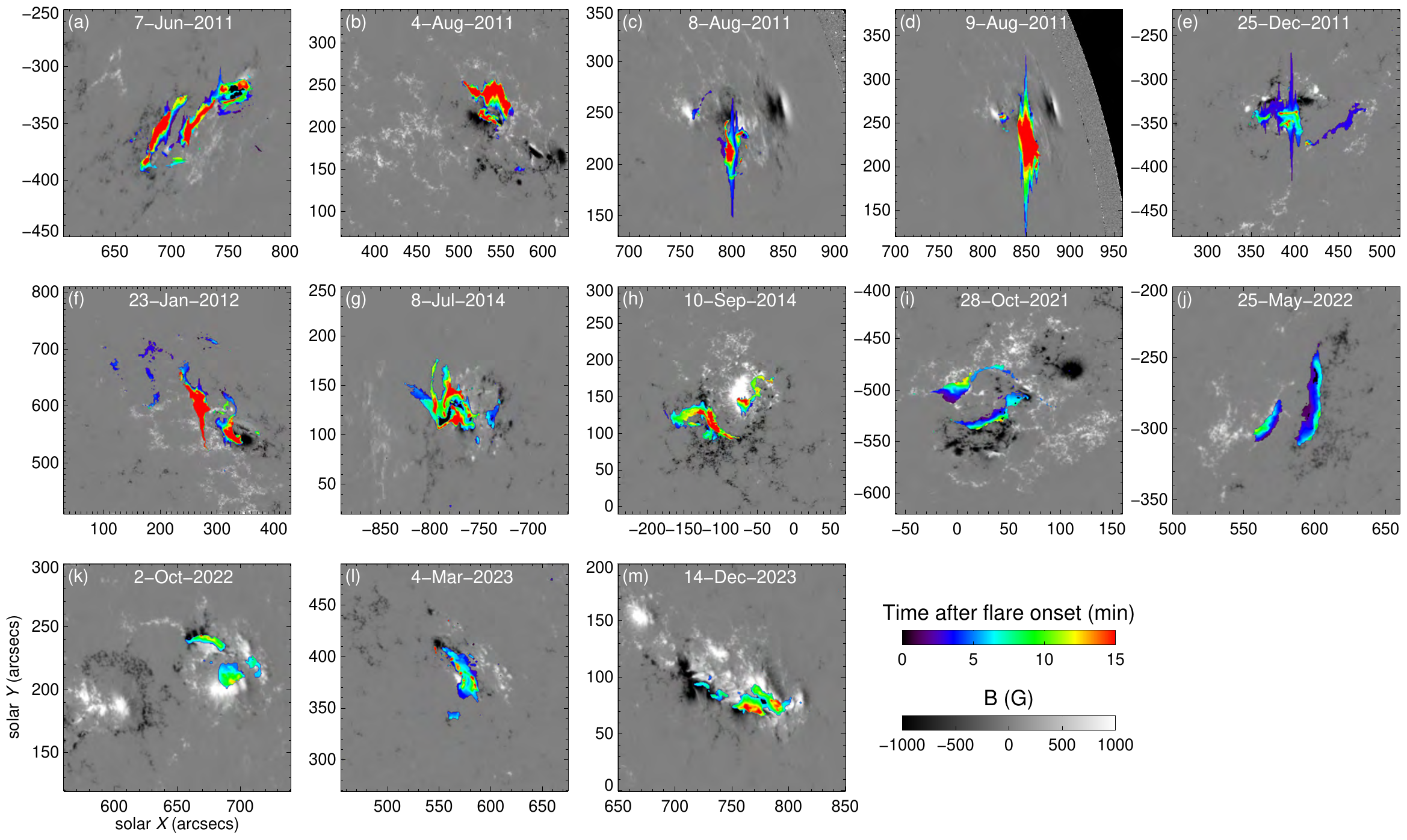}
\caption{Similar to Figure \ref{fig03} but for Group--\uppercase\expandafter{\romannumeral3} including 13 Moreton wave events. The magnetograms of 13 events display various morphology. Flare ribbons of all events are located close to the center of the active region.
} 
\label{fig05}
\end{figure*}

\begin{figure*}
\centering
\includegraphics[width=1.0\textwidth]{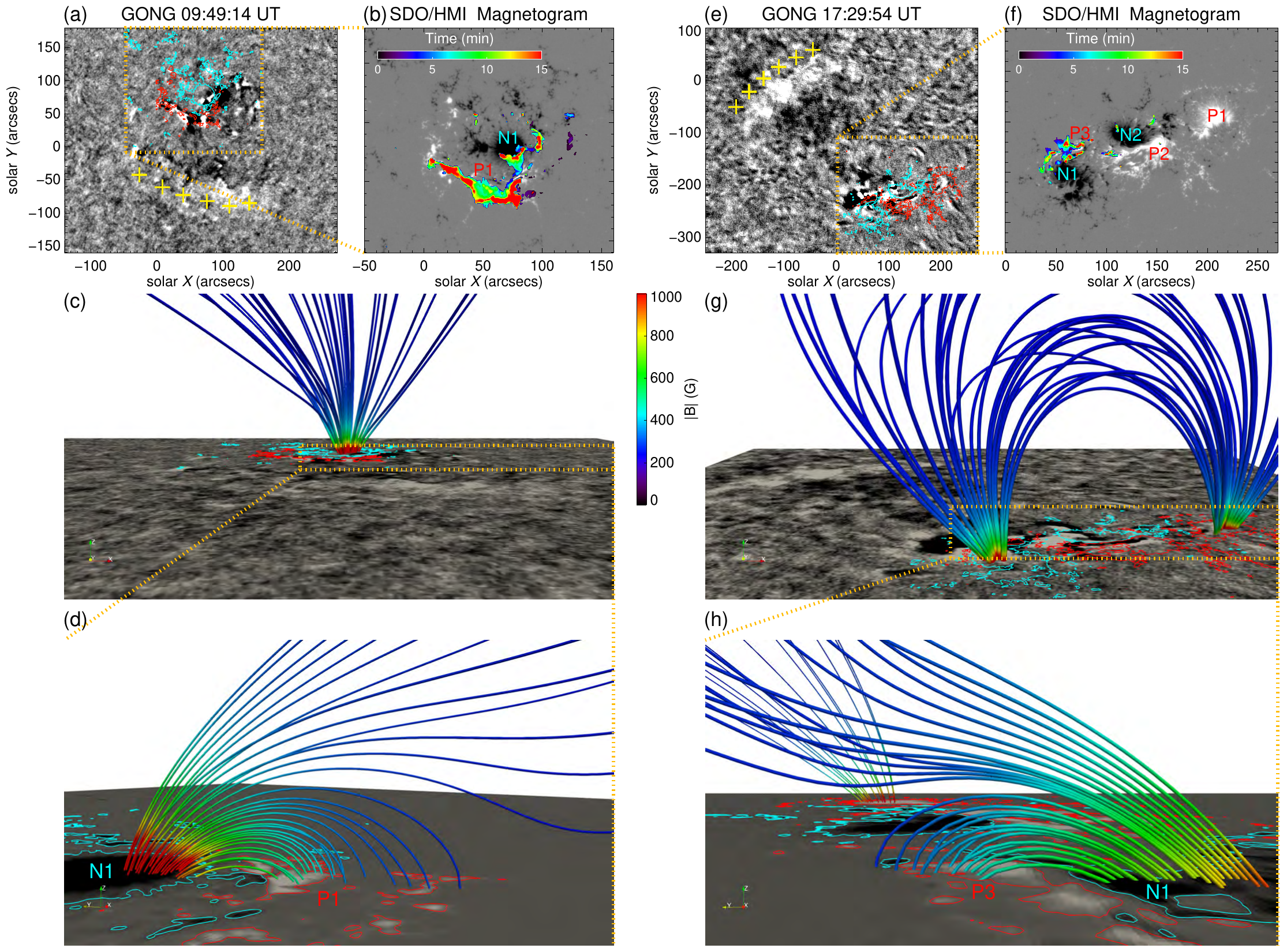}
\caption{Two main types of the magnetic configuration that generate Moreton waves: open fields and magnetic loops. (a) Running-difference GONG H$\alpha$ image of the Moreton wave occurred on 2011 October 1. Yellow plus signs depict the arc-shaped wave front. Cyan and red contours show the HMI line-of-sight magnetic field with $\pm$ 200 G. (b) The magnetogram includes a weaker positive polarity P1 in the south and a stronger negative polarity N1 in the north. Flare ribbons are superimposed on the magnetogram with the time in minutes from 09:40 UT indicated by the color bar. (c) A side view of selected 3D magnetic field lines along the $y$-axis. It shows the relative position between magnetic field lines and the source region that generates the Moreton wave. The bottom surface is the same as that in panel (a). (d) The magnetic field with a zoomed-in view of the region corresponding to the orange dotted box in panel (c) along the direction close to the $x$-axis. The field lines are inclined toward the southward ($y$-axis). (e)-(h) Similar to panels (a)--(d), but for the magnetic loop configuration that generates the Moreton wave on 2011 February 14. The magnetic field lines in panel (h) are inclined toward the northward ($y$-axis).
}
\label{fig06}
\end{figure*}

\begin{figure*}
\centering
\includegraphics[width=1.0\textwidth]{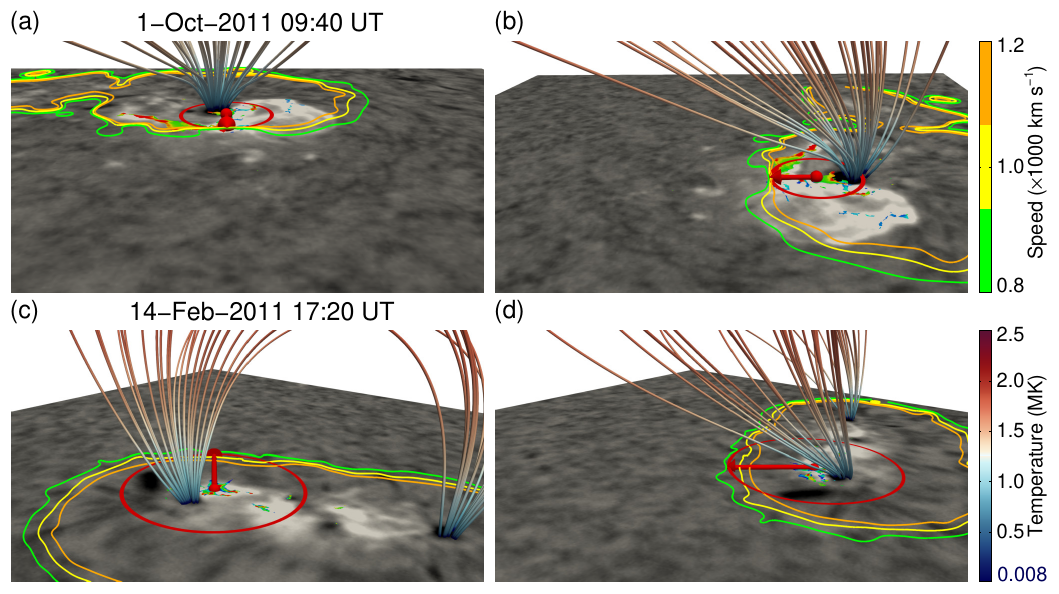}
\caption{Two groups of the magnetic configuration that generate Moreton waves.
(a) Moreton wave event occurred on 2011 October 1 with a side view along the direction of the wave propagation. Magnetic field lines are colored by the temperature. Orange, yellow and green contours represent calculated fast magnetoacoustic speed of 1200, 1000, and 800 km s$^{-1}$, respectively. The red sphere shows the center of the source eruptive region. The red arrow points to the direction where the speed of the horizontal fast-mode wave decreases the fastest, while the red circle is tangent to the green contour. The bottom boundary shows the GONG H$\alpha$ image at 09:40 UT when the flare occurs.
(b) A side view perpendicular to the direction of the wave propagation.
(c)--(d) Similar to panels (a)--(b), but for the Moreton wave event on 2011 February 14.
(An animation of this figure is available. Top row shows the evolution of the Moreton wave occurred on 2011 October 1st from 09:40 UT to 09:54 UT under inclined open fields while the bottom row shows the evolution of the Moreton wave occurred on 2011 February 14th from 17:20 UT to 17:34 UT under inclined magnetic loops. The duration of the animation is 3 seconds.)
}
\label{fig07}
\end{figure*}

\begin{figure*}
\centering
\includegraphics[width=1.0\textwidth]{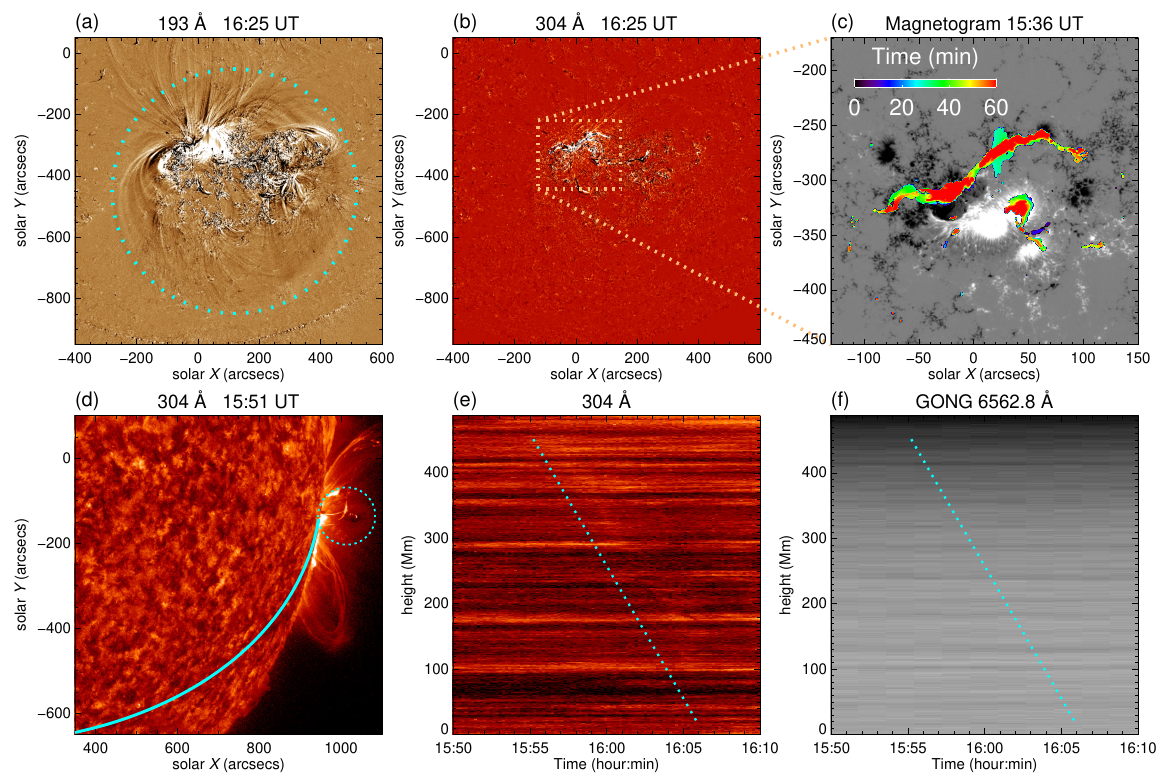}
\caption{Two X-class flares that did not produce observable Moreton waves occurred on 2012 July 12 ((a)--(c)), and 2017 September 10 ((d)--(f)), respectively.
(a) Running-difference image of AIA 193 {\AA} at 16:25 UT. The cyan dots depict the EUV wave front.
(b) AIA 304 {\AA} running-difference image at 16:25 UT, with no visible wave front.
(c) Photospheric magnetogram at 15:36 UT, overlaid with the evolution of flare ribbons at AIA 1600 {\AA} for 60 minutes from the flare onset time.
(d) AIA 304 {\AA} image shows the eruption snapshot at 15:51 UT. The cyan dots in the solar western limb mark the flux rope with a tear-drop shape. The cyan solid line points out the direction of the wave propagation.
(e) Time--distance diagram of AIA 304 {\AA} image displays the motion of the wave front along the cyan solid line in panel (d). The cyan dots indicate a speed of 677.16 $\pm$ 24.95 km s$^{-1}$.
(f) Similar to panel (e), but for the GONG H$\alpha$ time--distance diagram.
}
\label{fig08}
\end{figure*}

Owing to lack of multi-perspective observations for most events, it is impossible to estimate the inclination angles of these eruptions except two eruption events, which occurred on 2021 May 22 (Figure \ref{fig03}(m)) and 2011 August 4 (Figure \ref{fig05}(b)). With the dual perspective observations from SDO and STEREO, we determine the inclination angles of the two erupting filaments, which are $\sim$63$^\circ$ and $\sim$24$^\circ$, respectively. For the rest 64 events in our sample, although their eruption directions cannot be exactly determined, we have already demonstrated that their pre-eruptive magnetic configurations are inclined. \citet{2019Long} also suggested the high asymmetry of the magnetic field can cause the shock front to compress the chromosphere downwards significantly, which is consistent with our conclusion. These findings reinforce the requirement of non-radial or asymmetry for eruptions to generate Moreton waves suggested by \citet{2016Vrsnak}.

Base on the above analyses, we consider two primary and two secondary factors that contribute to the low occurrence rate of Moreton waves. One of the primary factors is the high inertia of the chromospheric plasma, which is several orders of magnitude denser than the corona, hence the CME piston-driven shock can easily perturb the corona, but not easily compress the chromosphere \citep{2002Chen,2005Chen}. Another primary factor is that most eruptions occur in the center of bipolar regions, beneath symmetric overlying fields. They do not have an inclined magnetic configuration where shock waves can compress the chromosphere favorably in the initial stage of the eruption.  According to the investigation of 904 filament eruptions from 2010 to 2014, \citet{2015McCauley} found that the proportion of non-radial eruptions is only 23\%. Their statistical results support our conclusion of ``inclined configuration''. As for the secondary factors, first, the overexpansion mentioned by \citet{2016Vrsnak} is an unusual phenomenon in the initial stages of CMEs, which is required to produce a shock strong enough to perturb the chromosphere. In addition, when the fast-mode wave pass through some local magnetic structures, it may lose the energy due to refraction or mode conversion. As a result, the fast-mode wave is not enough to compress the chromosphere and it is hard to discern Moreton waves in several observations as studied by \citet{2023Zheng}. A combination of multiple factors may explain the low occurrence rate of Moreton waves.

\acknowledgments
The authors are grateful to the referee for constructive comments and thank the GONG, CHASE, SDO/AIA and HMI consortia for supplying the data. This work was supported by the National Natural Science Foundation of China (12303061, 11973031, 12127901, 12333009), Shandong Natural Science Foundation of China (ZR2023QA074) and China Postdoctoral Science Foundation (2023T160385, 2022M711931). The CHASE mission is supported by CNSA. SDO is a mission of NASA’s Living With a Star Program.

\newpage

\bibliographystyle{aasjournal}
\bibliography{ms}

\newpage

\startlongtable
\begin{longrotatetable}
\begin{deluxetable*}{@{\extracolsep{1pt}} c c c c c c c c c c l}
\tablecolumns{12}
\centerwidetable
\centering
\tablecaption{List of 66 Moreton wave events}
\label{tab1}

\tablehead{
No & Date & Class\tablenotemark{1} & NOAA AR & Location\tablenotemark{2} & Start time & Peak time & End time\tablenotemark{3} & Group (label)\tablenotemark{4} & edge \tablenotemark{5} & References \tablenotemark{6}
}

\startdata
01  &	2011/02/14    &     M2.2	&    11158    &    S20W04	&    17:20    &    17:26    &    17:32    &    \uppercase\expandafter{\romannumeral2} (Fig. 4a)	& Y & \\
02  &	2011/03/12    &     M1.3	&    11166    &    N07W35	&    04:33    &    04:43    &    04:48    &    \uppercase\expandafter{\romannumeral2} (Fig. 4b)	& Y & \\
03  &	2011/06/07    &     M2.5	&    11226    &    S22W53	&    06:16    &    06:30    &    06:41    &    \uppercase\expandafter{\romannumeral3} (Fig. 5a)	& N & \\
04  &	2011/08/04    &     M9.3	&    11261    &    N16W38	&    03:41    &    03:45    &    03:57    &    \uppercase\expandafter{\romannumeral3} (Fig. 5b)	& N & \\
05  &	2011/08/08    &     M3.5	&    11263    &    N15W62	&    18:00    &    18:10    &    18:18    &    \uppercase\expandafter{\romannumeral3} (Fig. 5c)	& N & \\
06  &	2011/08/09    &     X6.9	&    11263    &    N14W69	&    07:48    &    08:05    &    08:08    &    \uppercase\expandafter{\romannumeral3} (Fig. 5d)	& N & \citet{2012Asai,2023Zheng} \\
07  &	2011/09/06    &     X2.1	&    11283    &    N14W18	&    22:12    &    22:20    &    22:24    &    \uppercase\expandafter{\romannumeral2} (Fig. 4c)	& Y & \\
08  &	2011/09/24    &     X1.9	&    11302    &    N13E61	&    09:21    &    09:40    &    09:48    &    \uppercase\expandafter{\romannumeral2} (Fig. 4d)	& Y & \\
09  &	2011/09/24    &     M3.0	&    11302    &    N12E42	&    19:09    &    19:21    &    19:41    &    \uppercase\expandafter{\romannumeral2} (Fig. 4e)	& Y & \\
10  &	2011/09/25    &     M3.7	&    11302    &    N13E44	&    15:26    &    15:33    &    15:38    &    \uppercase\expandafter{\romannumeral2} (Fig. 4f)	& Y & \\
11  &	2011/10/01    &     M1.2	&    11305    &    N09W04	&    09:40    &    09:59    &    10:17    &    \uppercase\expandafter{\romannumeral1} (Fig. 3a)	& Y & \\
12  &	2011/10/02    &     M3.9	&    11305    &    N10W14	&    00:37    &    00:50    &    00:59    &    \uppercase\expandafter{\romannumeral1} (Fig. 3b)	& Y & \\
13  &	2011/12/25    &     M4.0	&    11387    &    S22W26	&    18:11    &    18:16    &    18:20    &    \uppercase\expandafter{\romannumeral3} (Fig. 5e)	& N & \\
14  &	2011/12/26    &     M1.5	&    11387    &    S21W33	&    02:13    &    02:27    &    02:36    &    \uppercase\expandafter{\romannumeral2} (Fig. 4g)	& Y & \\
15  &	2012/01/23    &     M8.7	&    11402    &    N33W21	&    03:38    &    03:59    &    03:59    &    \uppercase\expandafter{\romannumeral3} (Fig. 5f)	& N & \\
16  &	2012/03/07    &     X1.3	&    11429    &    N15E26	&    01:05    &    01:14    &    01:23    &    \uppercase\expandafter{\romannumeral2} (Fig. 4h)	& Y & \\
17  &	2012/03/17    &     M1.3	&    11434    &    S20W25	&    20:32    &    20:39    &    20:42    &    \uppercase\expandafter{\romannumeral2} (Fig. 4i)	& Y & \\
18  &	2012/06/03    &     M3.3	&    11496    &    N16E38	&    17:48    &    17:55    &    17:57    &    \uppercase\expandafter{\romannumeral2} (Fig. 4j)	& Y & \citet{2015Admiranto} \\
19  &	2012/06/06    &     M2.1	&    11494    &    S18W04	&    19:54    &    20:06    &    20:13    &    \uppercase\expandafter{\romannumeral1} (Fig. 3c)	& Y & \\
20  &	2012/07/02    &     M3.8	&    11515    &    S17E00	&    19:59    &    20:07    &    20:13    &    \uppercase\expandafter{\romannumeral1} (Fig. 3d)	& Y & \\
21  &	2012/07/04    &     M4.6	&    11515    &    S16W28	&    22:03    &    22:09    &    22:15    &    \uppercase\expandafter{\romannumeral2} (Fig. 4k)	& Y & \\
22  &	2012/07/06    &     X1.1	&    11515    &    S13W59	&    23:01    &    23:08    &    23:14    &    \uppercase\expandafter{\romannumeral1} (Fig. 3e)	& Y & \citet{2015Admiranto} \\
23  &	2013/01/11    &     M1.2	&    11654    &    N05E36	&    08:43    &    09:11    &    09:17    &    \uppercase\expandafter{\romannumeral2} (Fig. 4l)	& Y & \\
24  &	2013/06/23    &     M2.9	&    11778    &    S15E62	&    20:48    &    20:56    &    20:59    &    \uppercase\expandafter{\romannumeral1} (Fig. 3f)	& Y & \\
25  &	2013/10/22    &     M4.2	&    11875    &    N04W00	&    21:15    &    21:20    &    21:22    &    \uppercase\expandafter{\romannumeral1} (Fig. 3g)	& Y & \\
26  &	2013/10/25    &     X2.1	&    11882    &    S06E69	&    14:51    &    15:03    &    15:12    &    \uppercase\expandafter{\romannumeral2} (Fig. 4m)	& Y & \\
27  &	2013/10/28    &     M4.4	&    11882    &    S06E28	&    15:07    &    15:15    &    15:21    &    \uppercase\expandafter{\romannumeral2} (Fig. 4n)	& Y & \\
28  &	2013/11/08    &     X1.1	&    11890    &    S13E13	&    04:20    &    04:26    &    04:29    &    \uppercase\expandafter{\romannumeral2} (Fig. 4o)	& Y & \\
29  &	2013/11/10    &     X1.1	&    11890    &    S13W13	&    05:08    &    05:14    &    05:18    &    \uppercase\expandafter{\romannumeral2} (Fig. 4p)	& Y & \\
30  &	2014/03/28    &     M2.0	&    12017    &    N10W20	&    19:04    &    19:18    &    19:27    &    \uppercase\expandafter{\romannumeral1} (Fig. 3h)	& Y & \citet{2019Long} \\
31  &	2014/03/28    &     M2.6	&    12017    &    N10W22	&    23:44    &    23:51    &    00:05    &    \uppercase\expandafter{\romannumeral1} (Fig. 3i)	& Y & \citet{2019Long} \\
32  &	2014/03/29    &     X1.0	&    12017    &    N10W32	&    17:35    &    17:48    &    17:54    &    \uppercase\expandafter{\romannumeral1} (Fig. 3j)	& Y & \citet{2016Francile,2019Cabezas} \\
33  &	2014/03/30    &     M2.1	&    12017    &    N10W42	&    11:48    &    11:55    &    12:02    &    \uppercase\expandafter{\romannumeral1} (Fig. 3k)	& Y & \citet{2019Long} \\
34  &	2014/07/08    &     M6.5	&    12113    &    N09E56	&    16:06    &    16:20    &    16:30    &    \uppercase\expandafter{\romannumeral3} (Fig. 5g)	& N & \\
35  &	2014/09/10    &     X1.6	&    12158    &    N11E05	&    17:21    &    17:45    &    17:45    &    \uppercase\expandafter{\romannumeral3} (Fig. 5h)	& N & \\
36  &	2015/11/04    &     M1.9	&    12445    &    N14W66	&    03:20    &    03:25    &    03:29    &    \uppercase\expandafter{\romannumeral2} (Fig. 4q)	& Y & \\
37  &	2015/11/04    &     M2.5	&    12445    &    N15W71	&    11:55    &    12:03    &    12:06    &    \uppercase\expandafter{\romannumeral2} (Fig. 4r)	& Y & \\
38  &	2016/04/18    &     M6.7	&    12529    &    N11W60	&    00:14    &    00:29    &    00:39    &    \uppercase\expandafter{\romannumeral2} (Fig. 4s)	& Y & \\
39  &	2017/04/01    &     M4.4	&    12644    &    N13W54	&    21:35    &    21:48    &    22:05    &    \uppercase\expandafter{\romannumeral2} (Fig. 4t)	& Y & \\
40  &	2017/04/02    &     M5.7	&    12644    &    N15W66	&    20:26    &    20:33    &    20:38    &    \uppercase\expandafter{\romannumeral2} (Fig. 4u)	& Y & \\
41  &	2017/07/14    &     M2.4	&    12665    &    S09W33	&    01:07    &    02:09    &    03:24    &    \uppercase\expandafter{\romannumeral2} (Fig. 4v)	& Y & \\
42  &	2019/05/06    &     M1.0	&    12740    &    N08E50	&    05:04    &    05:10    &    05:12    &    \uppercase\expandafter{\romannumeral1} (Fig. 3l)	& Y & \citet{2020Wang} \\
43  &	2021/05/22    &     M1.4	&    12824    &    N20E12	&    21:30    &    21:36    &    21:43    &    \uppercase\expandafter{\romannumeral1} (Fig. 3m)	& Y & \citet{2023Zheng} \\
44  &	2021/05/22    &     M1.1	&    12824    &    N18E18	&    17:03    &    17:11    &    17:16    &    \uppercase\expandafter{\romannumeral1} (Fig. 3n)	& Y & \\
45  &	2021/05/23    &     M1.1	&    12824    &    N20E05	&    11:00    &    11:08    &    11:14    &    \uppercase\expandafter{\romannumeral1} (Fig. 3o)	& Y & \\
46  &	2021/10/09    &     M1.6	&    12882    &    N18E08	&    06:19    &    06:38    &    06:53    &    \uppercase\expandafter{\romannumeral1} (Fig. 3p)	& Y & \\
47  &	2021/10/28    &     X1.0	&    12887    &    S28W01	&    15:17    &    15:35    &    15:48    &    \uppercase\expandafter{\romannumeral3} (Fig. 5i)	& N & \citet{2023Zheng} \\
48  &	2022/03/30    &     X1.3	&    12975    &    N13W31	&    17:21    &    17:37    &    17:46    &    \uppercase\expandafter{\romannumeral2} (Fig. 4w)	& Y & \\
49  &	2022/03/31    &     M9.6	&    12975    &    N12W47	&    18:17    &    18:35    &    18:45    &    \uppercase\expandafter{\romannumeral2} (Fig. 4x)	& Y & \\
50  &	2022/05/10    &     X1.5	&    13006    &    S29W04	&    13:50    &    13:55    &    13:59    &    \uppercase\expandafter{\romannumeral2} (Fig. 4y)	& Y & \\
51  &	2022/05/25    &     M1.3	&    13016    &    S19W41	&    18:09    &    18:24    &    18:43    &    \uppercase\expandafter{\romannumeral3} (Fig. 5j)	& N & \\
52  &	2022/07/14    &     M2.8	&    13058    &    N15E69	&    21:42    &    21:48    &    21:53    &    \uppercase\expandafter{\romannumeral2} (Fig. 4z)	& Y & \\
53  &	2022/08/15    &     M2.7	&    13078    &    S22W00	&    16:40    &    16:54    &    16:58    &    \uppercase\expandafter{\romannumeral1} (Fig. 3q)	& Y & \\
54  &	2022/08/16    &     M5.0	&    13078    &    S22W11	&    07:52    &    07:58    &    08:05    &    \uppercase\expandafter{\romannumeral1} (Fig. 3r)	& Y & \\
55  &	2022/10/02    &     X1.0	&    13110    &    N17W49	&    19:53    &    20:25    &    20:34    &    \uppercase\expandafter{\romannumeral3} (Fig. 5k)	& N & \\
56  &	2022/12/30    &     M1.4	&    13176    &    N18E08	&    15:24    &    15:28    &    15:32    &    \uppercase\expandafter{\romannumeral1} (Fig. 3s)	& Y & \\
57  &	2023/03/02    &     M3.8	&    13234    &    N21W64	&    21:05    &    21:16    &    21:25    &    \uppercase\expandafter{\romannumeral2} (Fig. 4aa)	& Y & \\
58  &	2023/03/04    &     M1.0	&    13243    &    N16W38	&    07:06    &    07:10    &    07:14    &    \uppercase\expandafter{\romannumeral3} (Fig. 5l)	& N & \\
59  &	2023/05/11    &     M2.2	&    13294    &    S06W41	&    08:47    &    09:01    &    09:11    &    \uppercase\expandafter{\romannumeral1} (Fig. 3t)	& Y & \\
60  &	2023/07/25    &     M1.6	&    13380    &    S10E40	&    21:08    &    21:16    &    21:24    &    \uppercase\expandafter{\romannumeral1} (Fig. 3u)	& Y & \\
61  &	2023/09/19    &     M4.0	&    13435    &    N07E45	&    20:01    &    20:14    &    20:21    &    \uppercase\expandafter{\romannumeral1} (Fig. 3v)	& Y & \\
62  &	2023/09/20    &     M8.2	&    13435    &    N07E35	&    14:11    &    14:19    &    14:25    &    \uppercase\expandafter{\romannumeral1} (Fig. 3w)	& Y & \\
63  &	2023/09/21    &     M8.7	&    13435    &    N07E22	&    12:42    &    12:54    &    13:02    &    \uppercase\expandafter{\romannumeral1} (Fig. 3x)	& Y & \\
64  &	2023/10/02    &     M1.9	&    13455    &    N18E69	&    12:35    &    12:46    &    12:58    &    \uppercase\expandafter{\romannumeral2} (Fig. 4ab)	& Y & \\
65  &	2023/11/28    &     M9.8	&    13500    &    S16W00	&    19:35    &    19:50    &    20:07    &    \uppercase\expandafter{\romannumeral2} (Fig. 4ac)	& Y & \\
66  &	2023/12/14    &     X2.8	&    13514    &    N04W53	&    16:47    &    17:02    &    17:12    &    \uppercase\expandafter{\romannumeral3} (Fig. 5m)	& N & \\
\enddata
\tablenotetext{1}{Magnitude is based on GOES classification.}
\tablenotetext{2}{The active region number and the location of the associated flare in columns 5 and 6.}
\tablenotetext{3}{The start, peak, and end time of the associated flares in columns 7,8 and 9.}
\tablenotetext{4}{The 66 events are divided into three groups according to magnetic field configuration. Group--\uppercase\expandafter{\romannumeral1} is related to open fields, Group--\uppercase\expandafter{\romannumeral2} is corresponding to magnetic loops, and the rest are classified into Group--\uppercase\expandafter{\romannumeral3}. Their corresponding labels are shown in Figures \ref{fig03}--\ref{fig05}.}
\tablenotetext{5}{Whether Moreton waves occur at the edge of the active region.}
\tablenotetext{6}{Literatures related to Moreton waves have been reported.}
\end{deluxetable*}
\end{longrotatetable}

\end{document}